\documentclass[3p,11pt,authoryear]{elsarticle}
\usepackage[T1]{fontenc}
\usepackage[utf8]{inputenc}
\title{A curvature-based criterion for harmonic circadian waveforms}

\author[1]{Yusuke Yamada}
\author[2]{Yutaro Kabata}
\author[3]{Ryoya Fukasaku}
\author[1]{Hiroshi Ito\corref{cor1}}
\cortext[cor1]{hito@design.kyushu-u.ac.jp}

\address[1]{Faculty of Design, Kyushu University, 4-9-1, Shiobaru, 815-8540, Fukuoka, Japan}
\address[2]{Graduate School of Science and Engineering, Kagoshima University, 1-21-24, Korimoto, 890-0065, Kagoshima, Japan}
\address[3]{Institute of Mathematics for Industry, Kyushu University, 744, Motooka, 819-0395, Fukuoka, Japan}
\ead{hito@design.kyushu-u.ac.jp}

\usepackage{bm}
\usepackage{pdfpages}
\usepackage{amsmath}
\usepackage{amsfonts}
\usepackage{graphicx}
\usepackage{caption}
\usepackage{float}
\usepackage{hyperref}
\graphicspath{{./}{Fig/}}

\hypersetup{
  hidelinks
}

\makeatletter
\renewcommand\subsection{\@startsection{subsection}{2}{\z@}%
  {-3.25ex\@plus -1ex \@minus -.2ex}%
  {1.5ex \@plus .2ex}%
  {\normalfont\normalsize\bfseries\sffamily}}
\makeatother
\begin{document}

%\flushbottom
%\maketitle

\begin{frontmatter}
\begin{abstract}
Experimental and theoretical studies of circadian rhythms have focused largely on the period, on mutants that alter it and on phase shifts, and this focus has driven the identification of clock genes and clarified how clocks entrain to light--dark cycles. The waveform of the oscillation itself, by contrast, has attracted little attention as an indicator of the properties of the underlying oscillator. To assess the waveform directly, we focus on whether the trajectory in the plane spanned by a variable and its time derivative possesses an inflection point, and we define an oscillation to be harmonic when no inflection point is present. Bioluminescence recordings from cyanobacteria and from the mammalian SCN were harmonic in this sense, as were most of the core clock components in mathematical models of the circadian clock. Numerical analysis of the Goodwin model, a minimal representation of the core circadian oscillator, yielded harmonic oscillation throughout. We confirmed this numerical trend semi-analytically using a piecewise-linearized Goodwin model. Because it evaluates the properties of a waveform without assuming a model structure, the approach we propose offers a new perspective on the waveform analysis of biological rhythms in general, not only circadian ones.
\end{abstract}

\end{frontmatter}
\thispagestyle{empty}
%\linenumbers

\section*{Introduction}
Circadian rhythms are physiological phenomena that recur with a period of approximately 24 hours and are observed across a wide range of organisms, from cyanobacteria to mammals. These rhythms are generated by self-sustained oscillators within cells, that is, by circadian clocks. Owing to this clock system, physiological circadian rhythms persist even under constant conditions. Clock genes have been identified by examining mutants that alter the free-running period, that is, the period under constant conditions \citep{NobelPrize:2017}. Another property of the circadian clock is that, by entraining to external cycles of daily variation such as light and temperature, it adjusts physiological functions such as the sleep--wake cycle and photosynthetic activity so that they occur at particular times of day \citep{dunlap2004chronobiology}. Mathematically, entrainment of circadian rhythms can be described in terms of a limit-cycle oscillator subject to periodic forcing. Entrainment has been analysed in depth using phase models, obtained by reducing the state of the oscillator to a single quantity, the phase \citep{winfree1967,pittendrigh1976}. The phase description has been particularly fruitful for analysing the collective behaviour of populations of oscillators \citep{kuramoto1984}.

However, the information that characterizes a circadian rhythm is not limited to its period and phase. The waveform of the rhythm is also linked to physiological function. For example, the external coincidence model has been proposed as a mechanism by which organisms sense photoperiod \citep{imaizumi_choosing_2025}. This model posits that a substance under circadian control opens a window of light sensitivity whenever it exceeds a threshold, and the model is thought to be employed in plants. As another example, the waveform of the glucocorticoid circadian rhythm governs the differentiation of precursor cells into fat cells, so that irregular feeding or sleep cycles, which distort this waveform, promote obesity \citep{bahrami-nejad_transcriptional_2018}.
It has also been shown that the waveform of the cAMP--PKA activity rhythm in the suprachiasmatic nucleus, rather than the circadian period itself, determines the length of the consolidated rest phase within a day \citep{tanaka2026}. Several recent mathematical analyses likewise point to a relationship between waveform and clock function. For instance, temperature compensation of the period is expected to be linked to the temperature dependence of the circadian waveform \citep{gibo2019,gibo2025}. It has also been shown that the closer the waveform is to a sinusoid, the more regular the circadian period becomes \citep{kaji2025}.

The idea of inferring the properties of an oscillator from the waveform itself has a long history. B\"unning focused on the waveform of the circadian clock and argued that the clock may possess the properties of both a harmonic and a relaxation oscillator \citep{bunning1973}. Takao Kondo, discussing the regularity of the cyanobacterial circadian clock and in particular of the KaiC phosphorylation rhythm, pointed out that the clock behaves much like a harmonic oscillator despite being a simple network of chemical reactions \citep{kondo_around_2021}. The terms harmonic oscillation and relaxation oscillation used in these discussions, however, originate in physics. In mechanics, a harmonic oscillator is a system with a linear restoring force arising near a potential minimum, a specific model described by $\ddot{x}=-kx$. Relaxation oscillation refers broadly to oscillations that combine a fast and a slow timescale; the spike-like action potential in neuroscience is one example \citep{izhikevich2007}. As models of relaxation oscillation, two-variable slow--fast systems of the form $\epsilon\dot{x}=f(x,y), \dot{y}=g(x,y)$, such as the van der Pol and FitzHugh--Nagumo equations, have been studied \citep{ginoux2009}. Yet no clear method has been proposed for quantifying, from an observed rhythmic waveform, how close that waveform is to a particular model.

Here we propose a method for classifying oscillatory waveforms from data without assuming a specific dynamical system. We define harmonic oscillation on the basis of the curvature of the trajectory of a periodic limit-cycle oscillation in the phase plane, and we then apply this classification to circadian clock systems. By examining the curvature of the trajectories that these systems trace in phase space, we show that experimental circadian data are close to harmonic oscillation. We further show that the Goodwin model, a mathematical model of a simple negative-feedback loop, exhibits harmonic oscillation throughout the parameter region examined.

\section*{Results}
\subsection*{Classification of oscillatory waveforms using curvature}
Let $x\colon I\to\mathbb{R}$ be a $C^3$ function. We call $x$ harmonic on $I$, or say that it belongs to the harmonic class, if its phase-plane trajectory $\Gamma(t)=(x(t),\dot{x}(t))$ has no inflection point on $I$. The term \textit{harmonic} is used here in a broader geometric sense, encompassing the classical harmonic oscillation $\ddot{x}=-kx$ with $k>0$ as a special case.

At each regular point of $\Gamma$, where $(\dot{x}(t),\ddot{x}(t))\neq(0,0)$, the curvature is
\begin{equation}
\kappa(t)
=
\frac{\dot{x}(t)\dddot{x}(t)-\ddot{x}(t)^2}
{\left(\dot{x}(t)^2+\ddot{x}(t)^2\right)^{3/2}}.
\label{eq:curvature}
\end{equation}
At an inflection point $t_0$, the curvature vanishes and hence satisfies $\kappa(t_0)=0$, or equivalently,
\begin{equation}
\dot{x}(t_0)\dddot{x}(t_0)
=
\ddot{x}(t_0)^2.
\label{eq:inflection}
\end{equation}
If $\ddot{x}(t_0)\neq0$, this relation implies that the Taylor expansion of $x$ through third order at $t_0$ agrees with that of a shifted exponential function of the form $\alpha\exp\!\left(\beta(t-t_0)\right)+\gamma$ for suitable constants $\alpha,\beta,\gamma\in\mathbb{R}$. Thus, in this local sense, an inflection point may be interpreted as a point at which $x(t)$ exhibits exponential-like behaviour. If $\ddot{x}(t_0)=0$, the corresponding third-order approximation is instead linear. For standard background on the differential geometry of plane curves, see \citep{doCarmo2016,BruceGiblin1992}.

We hereafter apply our classification to analyse a waveform of a self-sustained oscillator in a steady state, i.e., a limit cycle. As illustrations, we examine the curvature of several oscillators. A trivial example is the Stuart--Landau equation:
\begin{equation}
\begin{aligned}
\dot x =& a\,x - \omega\,y - (x^2 + y^2)\,(x - b\,y) \\
\dot y =& a\,y + \omega\,x - (x^2 + y^2)\,(y + b\,x),
\label{eq:stu}
\end{aligned}
\end{equation}
which, for $a>0$ and $\Omega:=\omega-ab\neq 0$, has the limit-cycle solution $x(t)=\sqrt{a}\cos\Omega t,\ y(t)=\sqrt{a}\sin\Omega t$ as $t\rightarrow \infty$ \citep{kuramoto1984}. Since the equation is equivariant under rotation of the $(x,y)$ plane, the choice of the observed variable amounts only to a shift of the time origin, and the resulting curvature is the same function of $t$ up to translation; it therefore suffices to consider $x$. The curvature is $-\frac{|\Omega|}{\sqrt{a}}\left[1+(\Omega^2-1)\cos^2\Omega t\right]^{-\frac{3}{2}}$, which is always negative. The system therefore always exhibits harmonic oscillation (Fig. 1). The Stuart--Landau equation is the normal form of a system close to a Hopf bifurcation point~\citep{kuramoto1984}. Exactly at the bifurcation point, its behaviour approaches that of the harmonic oscillator $\ddot{x}+x=0$ of classical mechanics. The term ``harmonic'', which we introduced in this paper, derives from this fact.

Next we turned to the FitzHugh--Nagumo model, a model of neuronal firing that tends to produce pulse-like oscillations:
% FitzHugh--Nagumo model
\begin{equation}
\begin{aligned}
\dot x =& a\Bigl(-\,y + x - \frac{x^3}{3}\Bigr) \\
\dot y =& x - b\,y + c.
\label{eq:fhn}
\end{aligned}
\end{equation}
For $a=5, b=0.4, c=0.6$, the limit-cycle trajectory of each variable possessed inflection points (Fig. 1); neither coordinate waveform was harmonic at these parameter values.

The curvature was negative at all times for the Stuart--Landau equation and over most of the cycle for the FitzHugh--Nagumo model. This is because differentiation advances the phase of each Fourier component by $\frac{\pi}{2}$, so that the trajectory in phase space rotates clockwise. 

\subsection*{Classification of experimental circadian data}
On the basis of the concept of harmonic class introduced above, we analysed experimental circadian data. The time series analysed here were collected manually from the published figures at an interval of 0.5~h, corresponding to 48 points per 24-h cycle. For the circadian rhythm of the prokaryote \textit{Synechococcus elongatus}, the standard approach is to monitor the activity of the circadian clock-regulated promoter by luciferase bioluminescence \citep{kondo1993}. Observation error must be removed from this bioluminescence time series before the curvature can be computed, and we therefore applied Gaussian smoothing (Methods). No inflection point appeared in the curvature computed from the denoised time series (Fig. 2).

As a second example, we examined the circadian rhythm of the suprachiasmatic nucleus, the central circadian pacemaker of the brain, in the mouse, an evolutionarily distant species. One approach to measuring mammalian circadian rhythms is to monitor the expression rhythm of the \textit{Per2} gene by bioluminescence, using a knock-in reporter in which luciferase is fused to the PER2 protein \citep{yoo2004}. Again, no inflection point was found in the Gaussian-smoothed bioluminescence rhythm of the mouse suprachiasmatic nucleus.
These two examples suggest that each component in circadian systems tends to be harmonic (Fig. 2 and Fig. S2).

\subsection*{Classification of waveforms from circadian clock models}
Many mathematical models of circadian rhythms based on experimentally identified molecular networks have been reported. Models based on the negative feedback of the \textit{per} gene, for instance, have been proposed \citep{goldbeter_model_1995,kurosawa2002,kim2012}. These models reproduce rhythms with a 24-hour period, but the properties of the waveform itself have not been examined. Given the tendency of experimental circadian data to be harmonic, we asked whether mathematical models exhibit it as well.

We first turned to the cyanobacterial circadian clock model of Sasai \citep{sasai2022}. That work developed a detailed mathematical model of the biochemical reactions underlying the KaiC phosphorylation rhythm, the core oscillator of cyanobacteria, in order to explore the mechanism of temperature compensation of the circadian period. Because Sasai's model describes the conformational changes and biochemical reactions of individual KaiC hexamers, its dynamics are stochastic. To remove the stochastic fluctuation we applied Gaussian smoothing, as we did to remove observation noise from the experimental data. No inflection point appeared in the curvature of the waveform of the mean phosphorylation level of the KaiC hexamer population (Fig. 3).

As another example, we examined the model of Kim and Forger \citep{kim2012}, a detailed mathematical model of the mammalian circadian clock that integrates positive and negative feedback loops and has 70 parameters and 180 variables. We computed the curvature of the resulting time series. Of the 180 variables, 27 showed no inflection point. The majority of the mRNA species of the core clock genes, including those of \textit{Per1}, \textit{Per2} and \textit{Bmals}, were among them (Fig. 3 and Fig. S3). Mathematical models of circadian rhythms can therefore also generate harmonic waveforms, at least for the central oscillator.

\subsection*{Harmonicity of the Goodwin model}
Both the mathematical models and the experimental data on circadian rhythms are likely to be harmonic. To clarify the reason, we turned to the Goodwin model, the simplest model commonly used to represent circadian rhythms (Fig. 4A) \citep{goodwin1965,kurosawa2002}:
% Goodwin model
\begin{equation}
\begin{aligned}
\dot{X} &= \frac{k_X}{K + Z^n} - \lambda_X X\\
\dot{Y} &= k_Y X - \lambda_Y Y\\
\dot{Z} &= k_Z Y - \lambda_Z Z.
\end{aligned}
\label{eq:goodwin}
\end{equation}
Here $X$ is the mRNA level of a clock gene, $Y$ the amount of the translated protein, and $Z$ the amount of protein that has been modified and translocated into the nucleus; $\lambda_X, \lambda_Y,\lambda_Z$ are the degradation rates of the respective molecules, and $n$ is a positive constant. In this context the Goodwin model represents negative feedback, in which a protein expressed through the central dogma represses the expression of its own gene.

We first examined the curvature of the limit-cycle trajectory in the $X$--$\dot{X}$ plane obtained by numerical integration under the conditions $n= 17, k_X= k_Y =k_Z =K= 1, \lambda_X= \lambda_Y= \lambda_Z=0.5$. No inflection point was present along the periodic orbit. To assess the dependence of harmonicity on the choice of variable, we also examined the curvature of the trajectories in the $Y$--$\dot{Y}$ and $Z$--$\dot{Z}$ planes. $Y$ and $Z$ were harmonic, as well as $X$ (Fig. 4B).

Next we turned to the parameter dependence. To reduce the number of parameters, we considered the model obtained by the rescaling $X = \frac{k_X}{Ks}x,
Y = \frac{k_Xk_Y}{Ks^2}y,
Z = K^{1/n}z,\tau = st$, where $s=\left(\frac{k_Xk_Yk_Z}{K^{1+1/n}}\right)^{1/3}$.
Note that this rescaling does not affect the presence or absence of inflection points. Assuming further that all degradation rates are equal,
we obtained the following model with two parameters.
\begin{equation}
\begin{aligned}
\frac{dx}{d\tau}
&=
\frac{1}{1+z^n}
-\lambda x,\\
\frac{dy}{d\tau}
&=
x-\lambda y,\\
\frac{dz}{d\tau}
&=
y-\lambda z,
\end{aligned}
\label{eq:goodwin2}
\end{equation}
where $\lambda=\frac{\lambda_X}{s}=\frac{\lambda_Y}{s}=\frac{\lambda_Z}{s}$. No inflection point was observed in the $x$--$\dot{x}$ trajectory for different values of $\lambda$; that is, the oscillatory waveform was harmonic (Fig. 4C).

To confirm that being harmonic is independent of the parameters, we examined trajectories over a wide range of Hill coefficients $n$ and values of $\lambda$, and no inflection point was observed for any parameter pair that produced a limit cycle (Fig. 4D). This suggests a link between harmonic class and the presence of a feedback loop. According to a linear stability analysis around the fixed point \citep{griffith1968}, system \eqref{eq:goodwin2} exhibits self-sustained oscillation in the range
\begin{equation}
\left(\frac{n}{n-8}\right)\left(\frac{8}{n-8}\right)^{\frac{1}{n}}\lambda^3<1,
    \label{eq:hopf}
\end{equation}
and undergoes a Hopf bifurcation at its boundary. When the parameters lie near the Hopf bifurcation point, the linear terms dominate and the waveform approaches a sinusoid; the trajectory is therefore expected to approach an elliptical orbit, as in the Stuart--Landau equation. Indeed, the maximum of the curvature moved further away from zero near the Hopf bifurcation point, that is, the trajectory was farther from developing an inflection point (Fig. 4D).

\subsection*{Harmonicity of the piecewise-linearized Goodwin model}
To confirm analytically the tendency of the Goodwin model to be harmonic found numerically in the preceding section, we turned to a simpler model. In the limit of Hill coefficient $n\to\infty$, Eq.~\eqref{eq:goodwin2} can be regarded as a piecewise-linear system with a switching boundary at $z=1$ (Fig. 5A):

\begin{equation}
\label{eq:hybrid_system}
\begin{aligned}
\begin{aligned}
\dot{x} &= 1 - \lambda x \\
\dot{y} &= x - \lambda y \\
\dot{z} &= y - \lambda z
\end{aligned}
\quad (z<1)
\qquad
\rightleftharpoons
\qquad
\begin{aligned}
\dot{x} &= - \lambda x \\
\dot{y} &= x - \lambda y \\
\dot{z} &= y - \lambda z
\end{aligned}
\quad (z\geq1).
\end{aligned}
\end{equation}
This model can be interpreted as one in which the synthesis of $x$ is repressed for $z\geq1$ and unrepressed for $z<1$. Each subsystem relaxes monotonically toward its own fixed point without oscillating, but the switching between them gives rise to a closed orbit (Fig. 5A).

To show analytically that the limit cycle of this piecewise-linear model has no inflection point in the $z$--$\dot{z}$ plane, we established the following two results. (i) We determined the range of $x_0, y_0$ for which the trajectory starting from a point $(x_0, y_0, 1)$ on the cross-section $z=1$ has no inflection point for any $t$. (ii) We showed that the limit-cycle trajectory passes through a point $(x_0, y_0, 1)$ within that range.Note that (i) provides a sufficient condition. Because the limit cycle oscillates across $z=1$, the time spent in each of the regions $z<1$ and $z\geq1$ is finite, and the trajectory samples the curvature only over that finite interval. Condition (i), by contrast, requires the absence of inflection points over all $t$ and therefore imposes a stronger constraint than necessary. Consequently, if (i) and (ii) hold then the limit cycle has no inflection point, but the converse does not necessarily follow.

For (i), the curvature can be obtained directly from the analytical solution of each linear subsystem (Methods). For (ii), we solved numerically the simultaneous equations that follow from the orbit being closed (Methods). For several values of $\lambda$, the limit-cycle trajectory satisfied the condition obtained in (i) (Fig. 5B).

The semi-analytical result of this section reduces the question of whether the limit cycle possesses an inflection point to the question of whether the passage point, recovered from the numerical solution of the simultaneous equations, satisfies the analytically determined condition \eqref{eq:cond}. The condition can be written down exactly as a function of $\lambda$, whereas the passage point was obtained numerically. Over the range of $\lambda$ examined, the passage point always satisfied the condition, consistent with the numerical results of the preceding section.

\section*{Discussion}
In this study we focused on the inflection points that appear along limit-cycle trajectories in order to classify circadian waveforms. Waveforms obtained from experimental circadian data and from numerical simulations of circadian clock models showed no inflection points; that is, they were confirmed to be harmonic oscillations. The Goodwin model in particular, which compactly represents the negative feedback of gene regulation, exhibited harmonic oscillation in the $x$--$\dot{x}$ plane throughout the $n$--$\lambda$ region we examined. In addition, for the piecewise-linear system obtained in the $n\to\infty$ limit, a semi-analytical approach showed that the limit cycle remains harmonic in the $z$--$\dot{z}$ plane over the range of $\lambda$ examined. These results suggest a link between the circadian clock being an oscillatory system built around a feedback loop and its generating waveforms without inflection points.

Attempts to extract useful information from oscillatory waveforms are not confined to chronobiology. In signal processing, for example, Fourier analysis is often applied. From the viewpoint of frequency analysis, a simple cascade can be regarded as a low-pass filter \citep{deronde2010}. Indeed, $\dot{x}=u(t)-\lambda x$ is a first-order system, and the Goodwin model, which contains two such stages, effectively incorporates the function of a low-pass filter. The attenuation of high-frequency components brings the waveform closer to a sinusoid and may thereby explain why inflection points become less likely to appear. The relationship between the frequency spectrum, a global measure, and the curvature, a local measure, remains an open problem beyond the scope of this report.

The harmonic oscillation treated in this study is a property of the waveform of a particular variable. It gives no guarantee about the properties of other variables. Indeed, in the Kim--Forger model the cytoplasmic and nuclear mRNA levels of \textit{Per1}, \textit{Per2}, \textit{Bmals} exhibited harmonic oscillation, whereas the cytoplasmic and nuclear mRNA levels of \textit{Cry1} showed inflection points (Fig. S3). As noted above, a simple signal transduction step $x_1 \rightarrow x_2$ removes high-frequency components, and downstream variables may therefore be less likely to possess inflection points. If the relationship between the topology of a molecular network and harmonic class were clarified, it would provide a clue for inferring topology from waveform.

We used Gaussian smoothing to analyse the experimental data. The choice of smoothing strength can affect the detection of inflection points. Here we adopted a smoothing strength sufficient to remove those inflection points judged to be non-essential. To analyse the large body of circadian waveform data accumulated in chronobiology from the viewpoint of inflection points, statistical methods for extracting the essential inflection points should be developed.

This study has focused on the waveform of the limit cycle to which the system converges after a long time. Even when the waveform is a harmonic oscillation, inflection points can arise along transient trajectories. Indeed, for the Goodwin model the curvature of the trajectory in the $z$--$\dot{z}$ plane for a system in state $(x,y,z)$ is
\begin{equation}
\kappa = \frac{y - x^2(1+z^n) - z\lambda - y^2(1+z^n)\lambda^2 + x(1+z^n)\lambda(y+z\lambda)}{(1+z^n)\left[(y-z\lambda)^2 + \left(x + \lambda(-2y+z\lambda)\right)^2\right]^{3/2}}.
\label{eq:goodwin_kappa}
\end{equation}
On the basis of this expression, the condition that the curvature vanishes defines an hourglass-shaped surface in the $x$--$y$--$z$ phase space (Fig. 6). The limit cycle does not intersect this surface. This geometric structure may underlie the parameter-independent harmonic oscillation of the limit cycle of the Goodwin model. How widely such a structure occurs in other circadian clock models and in real organisms will be an important clue to its generality.

The curvature-based analysis of oscillatory waveforms proposed in this paper requires no knowledge of the molecular mechanism of the system under study. It was for this reason that we were able to analyse experimental bioluminescence data from circadian rhythms. In other words, this was an attempt to discover harmonic-class oscillatory phenomena within chronobiology by using the curvature obtained from the set of derivatives~$(x, \dot{x}, \ddot{x}, \dddot{x})$~
in order to understand the observed data~$x(t)$. A closely related line of work is that of \citet{packard1980}, who proposed reconstructing the state space from an observed variable~$x(t)$~together with its time derivatives~$(x, \dot{x}, \ddot{x}, \dots)$~(rigorously justified by Takens' embedding theorem~\citep{takens1981}). The present study can also be regarded as one of the data-driven approaches that seek information about an underlying oscillatory system from experimental observations of its oscillatory dynamics.

Even when the underlying model is already known, the proposed method provides a novel characterization of oscillatory phenomena. For example, inflection points naturally partition a limit cycle into distinct segments. Just as the cell cycle is conventionally divided into the M, G1, S, and G2 phases, this approach may provide a meaningful partitioning of biological rhythm waveforms. A related approach is the Flow Curvature Method~\citep{ginoux2008, ginoux2009}, in which the slow manifold of a slow--fast system is characterized as the set of points where the flow curvature vanishes. These studies illustrate how concepts from differential geometry can be successfully applied to the analysis of dynamical systems.

The shape of the circadian waveform may contain more information than chronobiologists have so far expected. Our proposal is that harmonicity, defined through inflection points, can serve as one way of classifying circadian oscillatory waveforms. Just as changes in period contributed to the identification of many molecular networks, we hope that this measure of harmonicity will contribute to further elucidating the mechanisms of circadian rhythms.

\section*{Methods}
\subsection*{Numerical simulation}

To obtain the limit cycles of Eqs.~\eqref{eq:stu}, \eqref{eq:fhn} and \eqref{eq:goodwin2} numerically, we used the fourth-order Runge--Kutta method with $\Delta t = 10^{-3}$. We judged that an approximate limit-cycle solution had been obtained when the amplitudes $A_n, A_{n-1}$ (the difference between the maximum and the minimum within a cycle) of the $n$th and $(n-1)$th cycles satisfied $\left|A_n/A_{n-1} - 1\right| \le 10^{-4}$.
For the simulations of the KaiABC circadian clock model \citep{sasai2022} and of the mammalian circadian rhythm model \citep{kim2012} we used the programs provided by the respective papers. We solved the stochastic differential equations \eqref{eq:st_FN} and \eqref{eq:st_SL} numerically by the Euler--Maruyama method with a step of $\Delta\tau = 10^{-2}$ in the normalized time, corresponding to 100 steps per cycle. The resulting trajectories were subsampled to 50 points per cycle, and all subsequent curvature analyses were carried out on the subsampled series. 

\subsection*{Numerical computation of the curvature}
Computing the curvature requires numerical differentiation, for which we used the five-point central difference.
In addition, in the region where the curvature approaches zero, catastrophic cancellation occurred in double-precision arithmetic and the sign reversal of the curvature could not be determined reliably.
In such cases we performed the computation with 32 decimal digits of precision using the Python package \texttt{mpmath},
a library for arbitrary-precision floating-point arithmetic.
Inflection points were detected as sign changes of $\kappa$ between adjacent samples, with the crossing point located by linear interpolation.

\subsection*{Noise removal from experimental data}
\label{gauss}
Computing the curvature numerically from experimental data requires numerical differentiation, which is in general sensitive to noise \citep{chartrand2011}. Because the curvature requires a third derivative, repeated numerical differentiation amplifies the error. Observation noise must therefore be removed beforehand.

It is known that, for Gaussian smoothing applied to a trajectory, increasing the smoothing strength causes the pairwise annihilation of an inflection point at which the curvature along the trajectory approaches zero from above and one at which it approaches zero from below \citep{mokhtarian1992}. Only the non-essential inflection points originating from observation noise should be eliminated by Gaussian smoothing, whereas those originating from the waveform of the limit cycle carry information that should be retained. The two are distinguished by whether they are aperiodic or periodic.

We therefore first applied Gaussian smoothing at several strengths and observed the periodicity of the inflection points that disappeared.

We then considered, as synthetic data, the FitzHugh--Nagumo model with added noise
\begin{equation}
\begin{aligned}
\frac{1}{T}\dot{x} =& a\left(-y + x - \frac{x^{3}}{3}\right) + \sqrt{D}\xi_x(t)\\
\frac{1}{T}\dot{y} =& x - by + c + \sqrt{D}\xi_y(t),
\end{aligned}
\label{eq:st_FN}
\end{equation}
where $T$ is the period of the oscillator in the absence of noise, introduced in order to normalize the period to 1, and $\xi(t)$ is independent Gaussian noise satisfying $E[\xi(t)] = 0$ and $E\left[\xi(t)\xi(t')\right] = \delta(t - t')$, where $E[\cdot]$ denotes the expectation and $\delta(t)$ is the Dirac delta function.
Gaussian smoothing was then applied to the numerical solutions obtained from this model. The smoothing was implemented using the Python package \texttt{scipy.ndimage.gaussian\_filter1d}.
At high smoothing strength $\sigma$, inflection points present in every cycle were identified. We carried out the same analysis for the Stuart--Landau model in the presence of noise
\begin{equation}
\begin{aligned}
\frac{1}{T}\dot{x} =& ax - \omega y - (x^{2}+y^{2})(x-by) + \sqrt{D}\xi_x(t)\\
\frac{1}{T}\dot{y} =& ay + \omega x - (x^{2}+y^{2})(y+bx) + \sqrt{D}\xi_y(t),
\end{aligned}
\label{eq:st_SL}
\end{equation}
and found no periodically recurring inflection points even at values of $\sigma$ comparable to those used above (Fig. S1).
These results suggest that Gaussian smoothing makes it possible to extract the essential inflection points.

On this basis we applied Gaussian smoothing to the cyanobacterial and mouse SCN bioluminescence data and computed the curvature. In both cases, applying $\sigma =0.1$ eliminated the aperiodic inflection points (Fig. S2).

\subsection*{Set of trajectories without inflection points in the piecewise-linearized Goodwin model}
\label{step_i}

The analytical solution of Eq.~\eqref{eq:hybrid_system} under the initial condition $(x(0), y(0), z(0))=(x_0, y_0, z_0)$ is, for
$z<1$,
\begin{equation}
\begin{aligned}
x(t) &= \frac{1}{\lambda} + \left(x_0 - \frac{1}{\lambda}\right)e^{-\lambda t} \\[2mm]
y(t) &= \frac{1}{\lambda^2} + \left(y_0 - \frac{1}{\lambda^2}\right)e^{-\lambda t}
       + \left(x_0 - \frac{1}{\lambda}\right)t\,e^{-\lambda t} \\[2mm]
z(t) &= \frac{1}{\lambda^3} + \left(z_0 - \frac{1}{\lambda^3}\right)e^{-\lambda t}
       + \left(y_0 - \frac{1}{\lambda^2}\right)t\,e^{-\lambda t}
       + \left(x_0 - \frac{1}{\lambda}\right)\frac{t^2}{2}\,e^{-\lambda t},
\end{aligned} 
\label{eq:analyticsol1}
\end{equation}

and, for $z\geq 1$,
\begin{equation}
\begin{aligned}
x(t) &= x_0\,e^{-\lambda t} \\[2mm]
y(t) &= \left(y_0 + x_0 t\right)e^{-\lambda t} \\[2mm]
z(t) &= \left(z_0 + y_0 t + \frac{x_0 t^2}{2}\right)e^{-\lambda t}.
\end{aligned}
\label{eq:analyticsol2}
\end{equation}

For $(x(0),y(0),z(0))=(x_0, y_0, 1)$, the numerator $\kappa_N(t)$ of the curvature $\kappa(t)$ of the $z$--$\dot{z}$ trajectory computed from these analytical solutions is
\begin{equation}
\kappa_N(t)=\left\{
\begin{aligned}
  &\frac{1}{2}e^{-2t\lambda}\Bigl[-t^2 - 2\lambda - 2y_0^2\lambda^2 + 2y_0(1+t\lambda) \\
  &\hphantom{\frac{1}{2}e^{-2t\lambda}\Bigl[}
    - x_0^2\left(2-2t\lambda+t^2\lambda^2\right) \\
  &\hphantom{\frac{1}{2}e^{-2t\lambda}\Bigl[}
    + 2x_0\left(t^2\lambda + \lambda(y_0+\lambda) - t\left(1+y_0\lambda^2\right)\right)\Bigr]
    && (z<1) \\[8pt]
  &\frac{1}{2}e^{-2t\lambda}\Bigl[-2y_0^2\lambda^2 + 2x_0\lambda\left(y_0+\lambda-y_0t\lambda\right) \\
  &\hphantom{\frac{1}{2}e^{-2t\lambda}\Bigl[}
    - x_0^2\left(2-2t\lambda+t^2\lambda^2\right)\Bigr]
    && (z\geq1).
\end{aligned}\right.
\end{equation}

Consequently, the condition that there be no $t$ at which $\kappa_N(t)=0$ is
\begin{equation}
\begin{cases}
-4(-1+x_0\lambda)^2\!\left(x_0^2-2y_0+2\lambda-2x_0\lambda^2+y_0^2\lambda^2\right)<0 & (z<1)\\
-4x_0^2\lambda^2\!\left(x_0^2-2x_0\lambda^2+y_0^2\lambda^2\right)<0 & (z\geq1).
\end{cases}
\end{equation}
Adding a condition $0<\lambda<1$ gives
\begin{equation}
\begin{cases}
  \left(x_0-\lambda^2\right)^2+\lambda^2\left(y_0-\dfrac{1}{\lambda^2}\right)^2>\left(\lambda^2-\dfrac{1}{\lambda}\right)^2 &
  (z<1) \\[4mm]
  \left(x_0-\lambda^2\right)^2+\lambda^2 y_0^2>\lambda^4 &
  (z\geq1).
\end{cases}
\label{eq:cond}
\end{equation}
The sets of $(x_0,y_0)$ satisfying \eqref{eq:cond} for different values of $\lambda$ are shown in Fig. 5B.

\subsection*{Limit-cycle solution of the piecewise-linearized Goodwin model}
\label{step_ii}
\begin{sloppypar}
Figure 5B requires the limit-cycle solution of the piecewise-linear system, Eq.~\eqref{eq:hybrid_system}, of the Goodwin model. Let $(x,y,z)=(\alpha_x, \alpha_y, 1)$ denote the intersection of the trajectory with the cross-section when this limit-cycle solution crosses $z=1$ with $\dot{z}<0$. Let $\alpha_t$ denote the time taken to traverse the region $z<1$; since the solution lies on a limit cycle, the system returns to $z=1$ after a time $\alpha_t$ and crosses it with $\dot{z}>0$. Let $(x,y,z)=(\beta_x, \beta_y, 1)$ denote this second intersection.

From the analytical solution \eqref{eq:analyticsol1}, $\beta_x$ and $\beta_y$ satisfy

\begin{equation}
\begin{aligned}
\beta_x &= \frac{1}{\lambda} + \left(\alpha_x - \frac{1}{\lambda}\right) e^{-\lambda \alpha_t} \\
\beta_y &= \frac{1}{\lambda^2} + \left(\alpha_y - \frac{1}{\lambda^2}\right) e^{-\lambda \alpha_t}
         + \left(\alpha_x - \frac{1}{\lambda}\right) \alpha_t\, e^{-\lambda \alpha_t} \\
1 &= \frac{1}{\lambda^3} + \left(1 - \frac{1}{\lambda^3}\right) e^{-\lambda \alpha_t}
   + \left(\alpha_y - \frac{1}{\lambda^2}\right) \alpha_t\, e^{-\lambda \alpha_t}
   + \left(\alpha_x - \frac{1}{\lambda}\right) \frac{\alpha_t^2}{2}\, e^{-\lambda \alpha_t}.
\end{aligned}
\label{eq:alpha_to_beta}
\end{equation}

Similarly, since the orbit is a limit cycle, the point reached after passing through $(x,y,z)=(\beta_x, \beta_y, 1)$ and returning to the plane $z=1$ is $(\alpha_x, \alpha_y, 1)$. Let $\beta_t$ denote the time taken to traverse the region $z\geq1$. The analytical solution \eqref{eq:analyticsol2} for $z\geq1$ of Eq.~\eqref{eq:hybrid_system} then gives the following conditions on $\alpha_x$ and $\alpha_y$.
\begin{equation}
\begin{aligned}
\alpha_x &= \beta_x\, e^{-\lambda \beta_t} \\
\alpha_y &= \left(\beta_y + \beta_x \beta_t\right) e^{-\lambda \beta_t} \\
1 &= \left(1 + \beta_y \beta_t + \beta_x \frac{\beta_t^2}{2}\right) e^{-\lambda \beta_t}.
\end{aligned}
\label{eq:beta_to_alpha}
\end{equation}

Equations \eqref{eq:alpha_to_beta} and \eqref{eq:beta_to_alpha} give six simultaneous equations for the unknowns $(\alpha_x, \alpha_y, \beta_x, \beta_y, \alpha_t, \beta_t)$. Eliminating $\alpha_x, \alpha_y, \beta_x, \beta_y$ reduces these to two simultaneous equations involving only the residence times $(\alpha_t, \beta_t)$:

\begin{equation}
\label{eq:hybrid_reduced_time}
\begin{aligned}
G_1(\alpha_t,\beta_t;\lambda) &= 0\\
G_2(\alpha_t,\beta_t;\lambda) &= 0,
\end{aligned}
\end{equation}
where, writing $W=e^{(\alpha_t+\beta_t)\lambda}-1$,
\begin{equation}
\begin{aligned}
G_1 &= e^{-\alpha_t\lambda}\Bigl[\,
 -2+2\lambda^{3}+2e^{(3\alpha_t+2\beta_t)\lambda}
\\
&\hphantom{{}=e^{-\alpha_t\lambda}\Bigl[\,}
 + \left(2-2\alpha_t\lambda+\alpha_t^{2}\lambda^{2}\right)e^{\alpha_t\lambda}
\\
&\hphantom{{}=e^{-\alpha_t\lambda}\Bigl[\,}
 + \left(4+2\alpha_t\lambda-\alpha_t^{2}\lambda^{2}-2\alpha_t\beta_t\lambda^{2}-4\lambda^{3}\right)
   e^{(\alpha_t+\beta_t)\lambda}
\\
&\hphantom{{}=e^{-\alpha_t\lambda}\Bigl[\,}
 + \left(-4+2\alpha_t\lambda+\alpha_t^{2}\lambda^{2}+2\alpha_t\beta_t\lambda^{2}\right)
   e^{(2\alpha_t+\beta_t)\lambda}
\\
&\hphantom{{}=e^{-\alpha_t\lambda}\Bigl[\,}
 - \left(2+2\alpha_t\lambda+\alpha_t^{2}\lambda^{2}-2\lambda^{3}\right)
   e^{2(\alpha_t+\beta_t)\lambda}\,\Bigr]
\\
&\quad -2\lambda^{3}W^{2}
\\[4mm]
G_2 &= \beta_t\Bigl[\,
 2-\beta_t\lambda
\\
&\hphantom{{}=\beta_t\Bigl[\,}
 + \left(-2+2\alpha_t\lambda+\beta_t\lambda\right)e^{\alpha_t\lambda}
\\
&\hphantom{{}=\beta_t\Bigl[\,}
 - \left(2+2\alpha_t\lambda+\beta_t\lambda\right)e^{(\alpha_t+\beta_t)\lambda}
\\
&\hphantom{{}=\beta_t\Bigl[\,}
 + \left(2+\beta_t\lambda\right)e^{(2\alpha_t+\beta_t)\lambda}\,\Bigr]
\\
&\quad -2\lambda^{2}\left(1-e^{-\beta_t\lambda}\right)W^{2}.
\end{aligned}
\end{equation}
\end{sloppypar}

\section*{Acknowledgments}
We are grateful to the late Takao Kondo for inspiring this study, and to the Education and Research Center for Mathematical and Data Science, Kyushu University, for fostering the research community from which this study emerged. This work was supported by the Japan Society for the Promotion of Science (JSPS) KAKENHI to H.I. (JP23H04475, JP25H02463), Y.K. (JP25H01485), R.F. (JP25H01482), AMED CREST to H.I. (JP24gm2010005), and JST SPRING to Y.Y. (JPMJSP2136). The funders had no role in the study design, data collection and analysis, decision to publish, or manuscript preparation. Claude Sonnet and Opus (Anthropic) were used to improve the manuscript's grammar and readability, and the authors reviewed and edited all AI-generated suggestions, taking full responsibility for the final content.

\section*{Author contributions}
Y.Y., Y.K., and H.I. designed the work and wrote the manuscript. Y.Y. performed the numerical simulations. Y.Y., Y.K., R.F., and H.I. analysed the data. Y.K. and H.I. provided guidance and supervision throughout the project.

\section*{Data availability}
All data are available upon reasonable request to the corresponding author.

\section*{Code availability}
All code is available on GitHub at \url{https://github.com/hitolab/harmonic}.

\section*{Figure legends}
\noindent\textbf{Fig. 1 Inflection points of trajectories in the Stuart--Landau and FitzHugh--Nagumo systems.} The top row shows the time course of each model, the middle row the trajectory in the space of the variable and its time derivative, and the bottom row the time course of the curvature $\kappa(t)$ along the trajectory. Parameters were set to $a = 4, b = 0.3, \omega = 0.4$ for the Stuart--Landau model and $a = 5, b = 0.4, c = 0.6$ for the FitzHugh--Nagumo model.\\

\noindent\textbf{Fig. 2 Curvature of trajectories from experimental circadian data.} Bioluminescence from the \textit{psbAI} promoter reporter in the cyanobacterium \textit{Synechococcus elongatus} and from the PER2::LUC fusion protein in the mouse SCN was recorded under constant conditions. The bioluminescence level, the trajectory in the space of the value and its time derivative, and the curvature are shown. Because the waveform is prone to distortion by smoothing near the beginning and the end of the recording, the time series before the first peak and after the last peak were excluded from the analysis.\\

\noindent\textbf{Fig. 3 Curvature of trajectories from circadian clock models.} Time courses of the cytoplasmic \textit{Per1} mRNA in the Kim--Forger model and of the KaiC phosphorylation ratio in the Sasai model, together with the trajectories in the space of the value and its time derivative and the curvature, are shown. The stochastic component of the Sasai model trajectory was removed by Gaussian smoothing with $\sigma=0.1$.\\

\noindent\textbf{Fig. 4 The Goodwin model is harmonic regardless of the parameter values.} (A) Schematic of the Goodwin model, representing the negative feedback formed by three biomolecules. (B) Time course, trajectory and curvature of $x$, $y$, and $z$. $n=17, \lambda=0.5$. (C) Time course, trajectory and curvature of $x$ for different values of $\lambda$. $n=17, \lambda=0.2, 0.5, 0.8$. (D) Maximum of $\kappa S$ over the limit cycle in the parameter space $n$--$\lambda$, where $S$ is the perimeter of the limit cycle in the phase plane. To make the values comparable across conditions, time was rescaled by the period of each limit cycle and the curvature was multiplied by $S$, so that both the period and the perimeter are equal to unity. The yellow line marks the Hopf bifurcation, where the left-hand side of Eq.~\eqref{eq:hopf} equals unity; in the black region the system does not oscillate and converges to a fixed point. Values outside the colour-bar range are shown in the limiting colours.\\

\noindent\textbf{Fig. 5 Initial values on $z=1$ for which the trajectory has no inflection point in the piecewise-linear model.} (A) The piecewise-linear systems defined separately for $z<1$ and $z\geq 1$, obtained in the limit $n\rightarrow \infty$ of the Goodwin model. (B) The set on the cross-section $z=1$ for which the trajectory has no inflection point. Crosses mark the passage points of the numerically obtained limit cycle.\\

\noindent\textbf{Fig. 6 Set of inflection points in the Goodwin model.} The magenta curve and the white circle indicate the limit-cycle solution and the unstable fixed point of the system. The cyan surface is the set of $(x, y, z)$ satisfying $\dddot{z}\dot{z}-\ddot{z}^2=0$. $\lambda=0.6, n=12$.\\

\bibliographystyle{elsarticle-harv}
\bibliography{cas-refs}

\ifdefined\nolinenumbers\nolinenumbers\fi
\includepdf[pages=-]{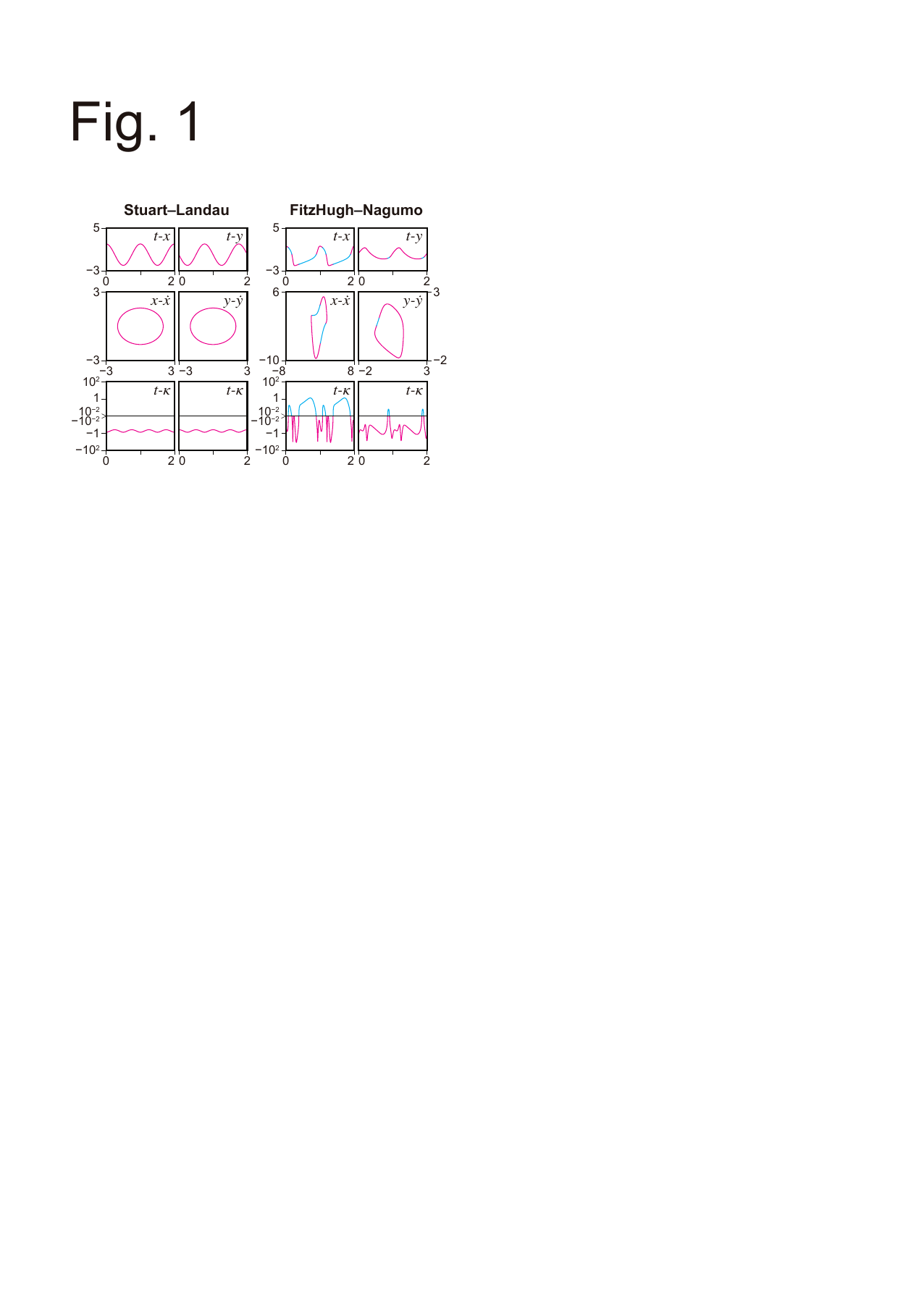}
\includepdf[pages=-]{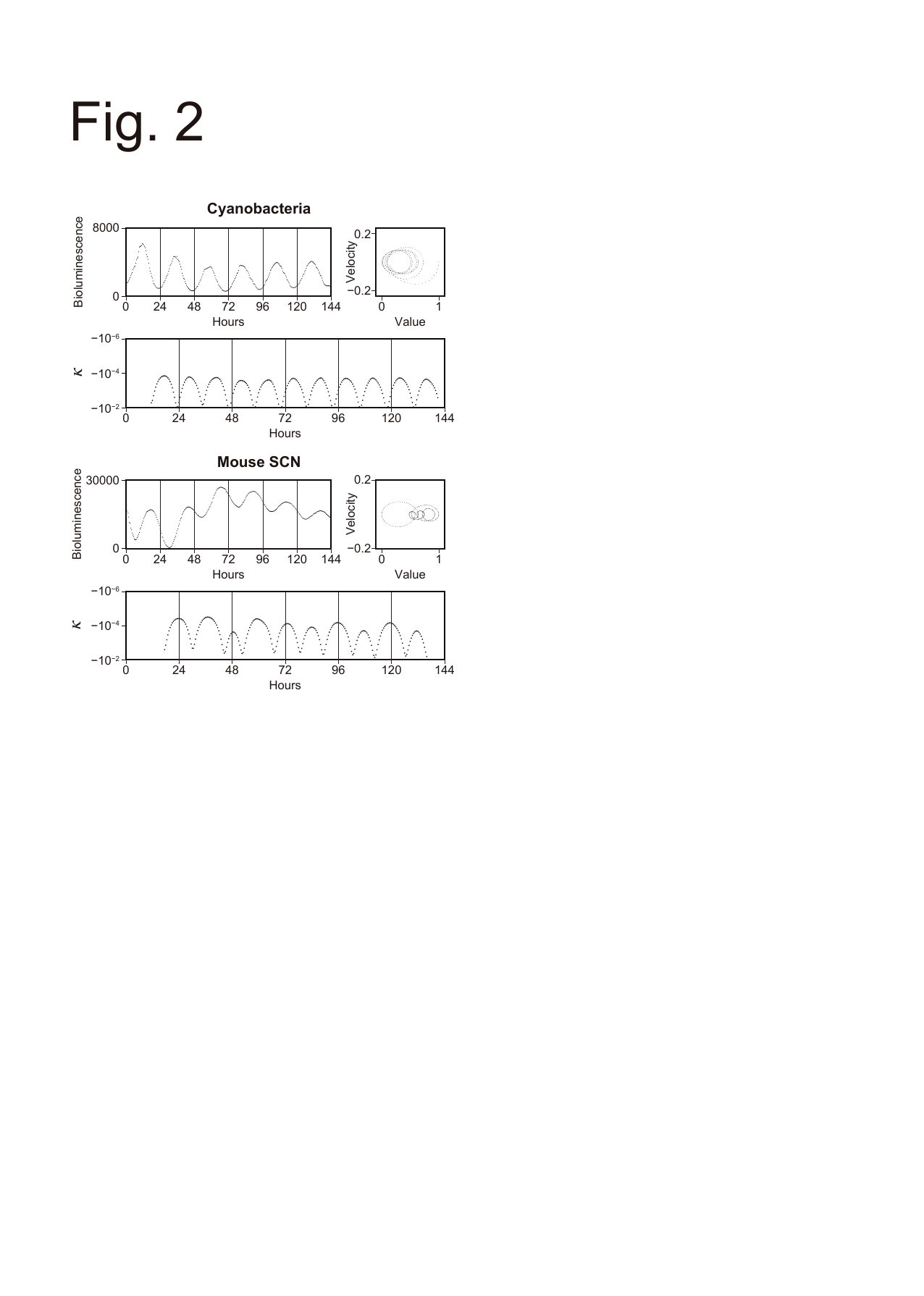}
\includepdf[pages=-]{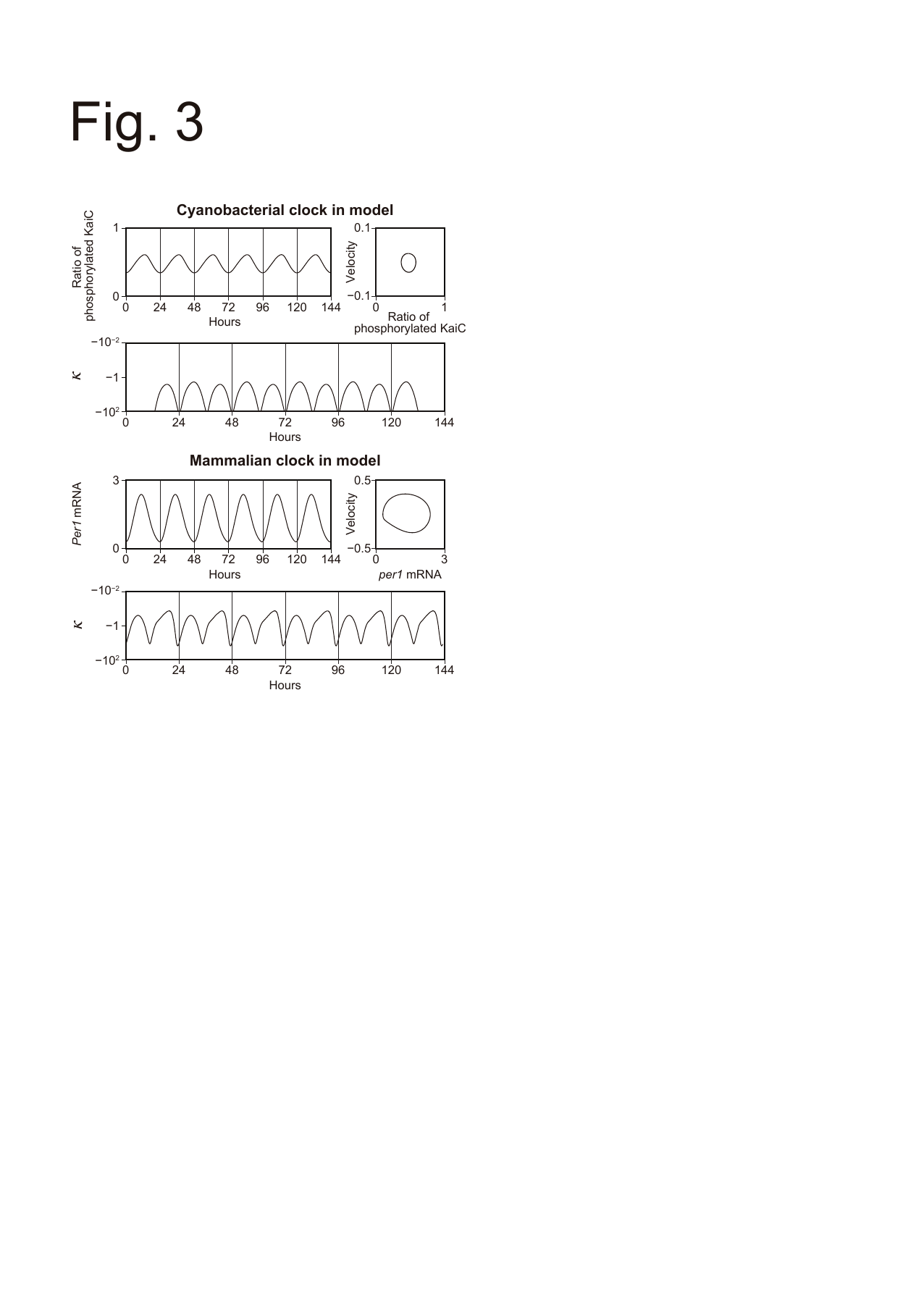}
\includepdf[pages=-]{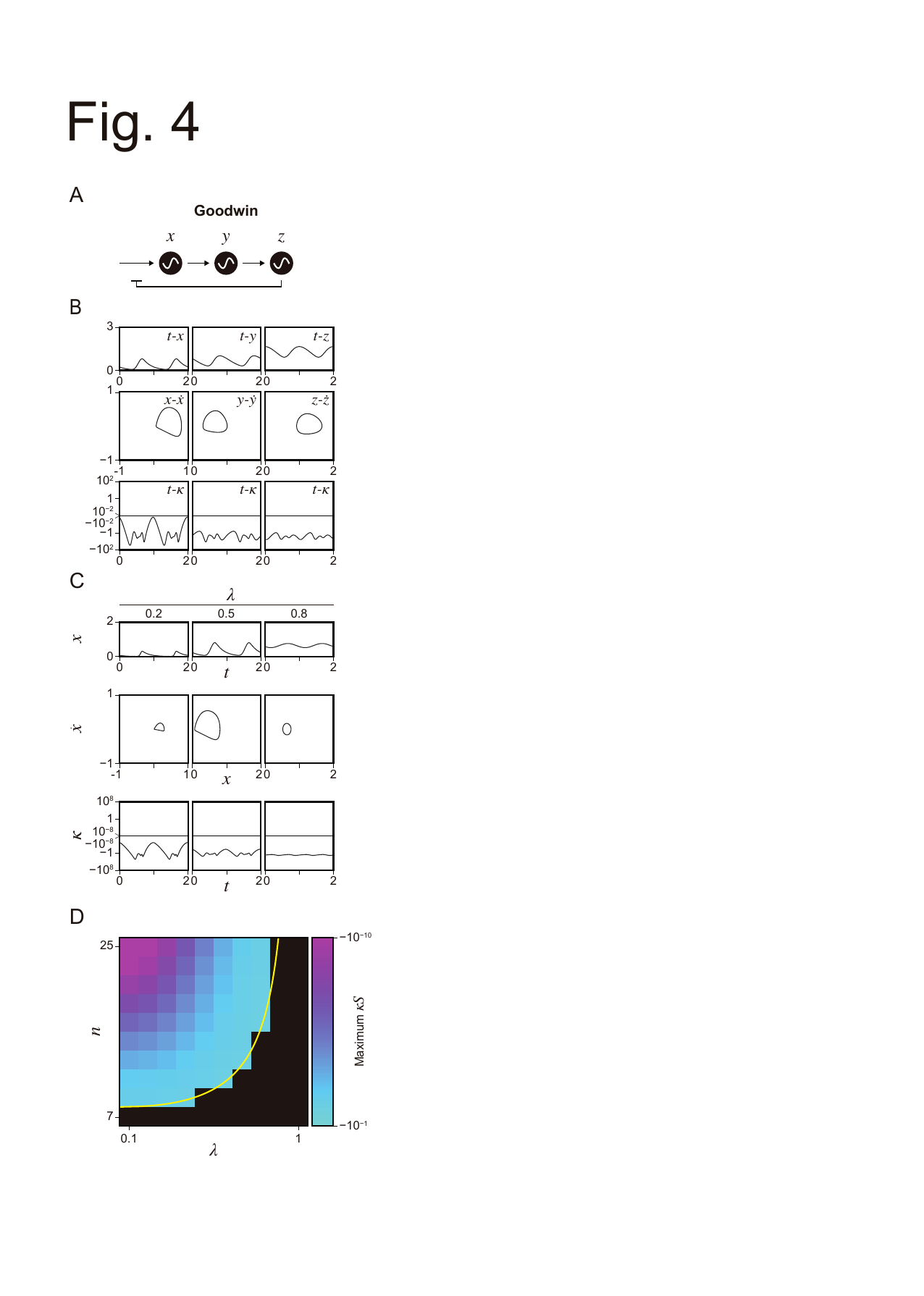}
\includepdf[pages=-]{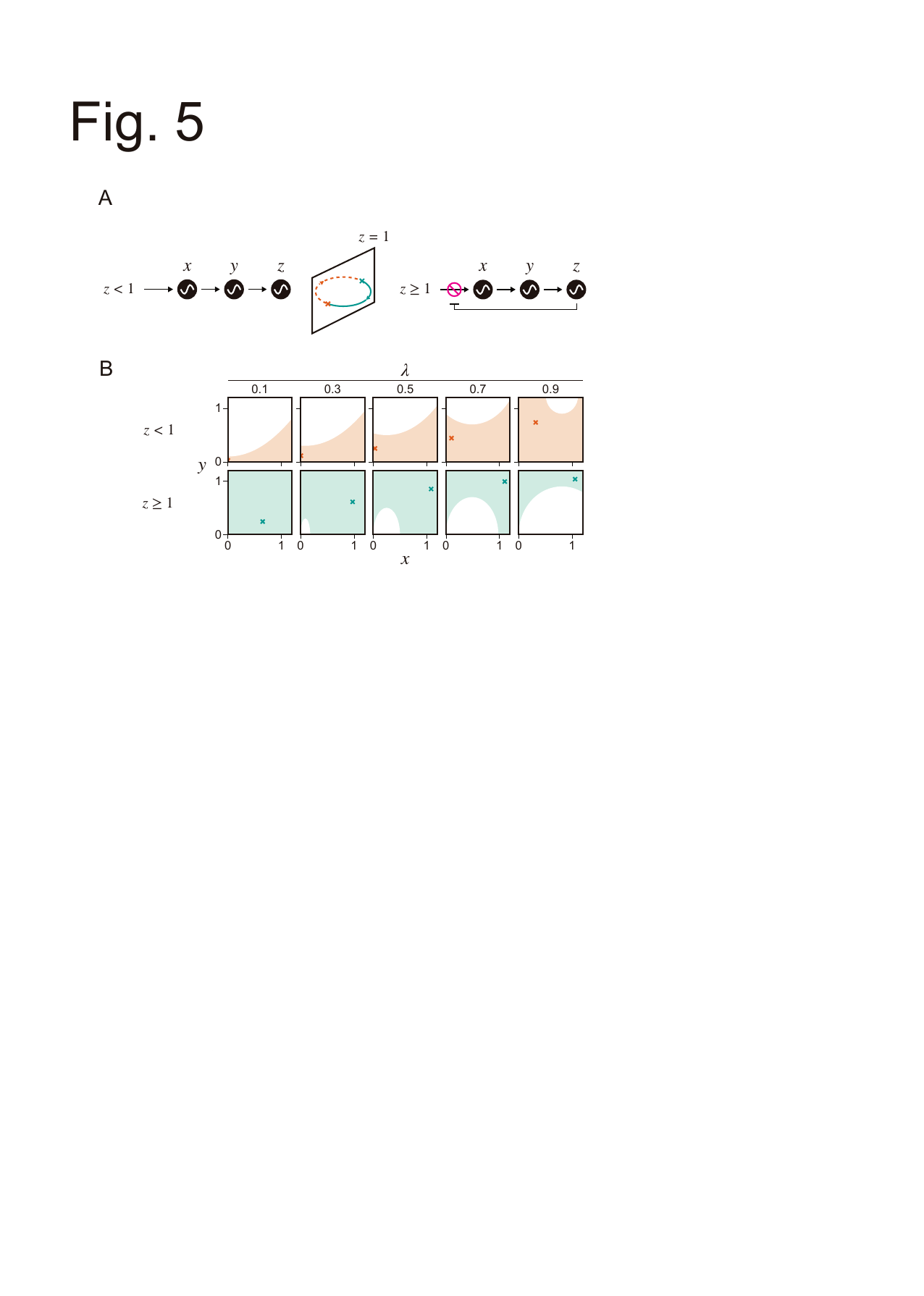}
\includepdf[pages=-]{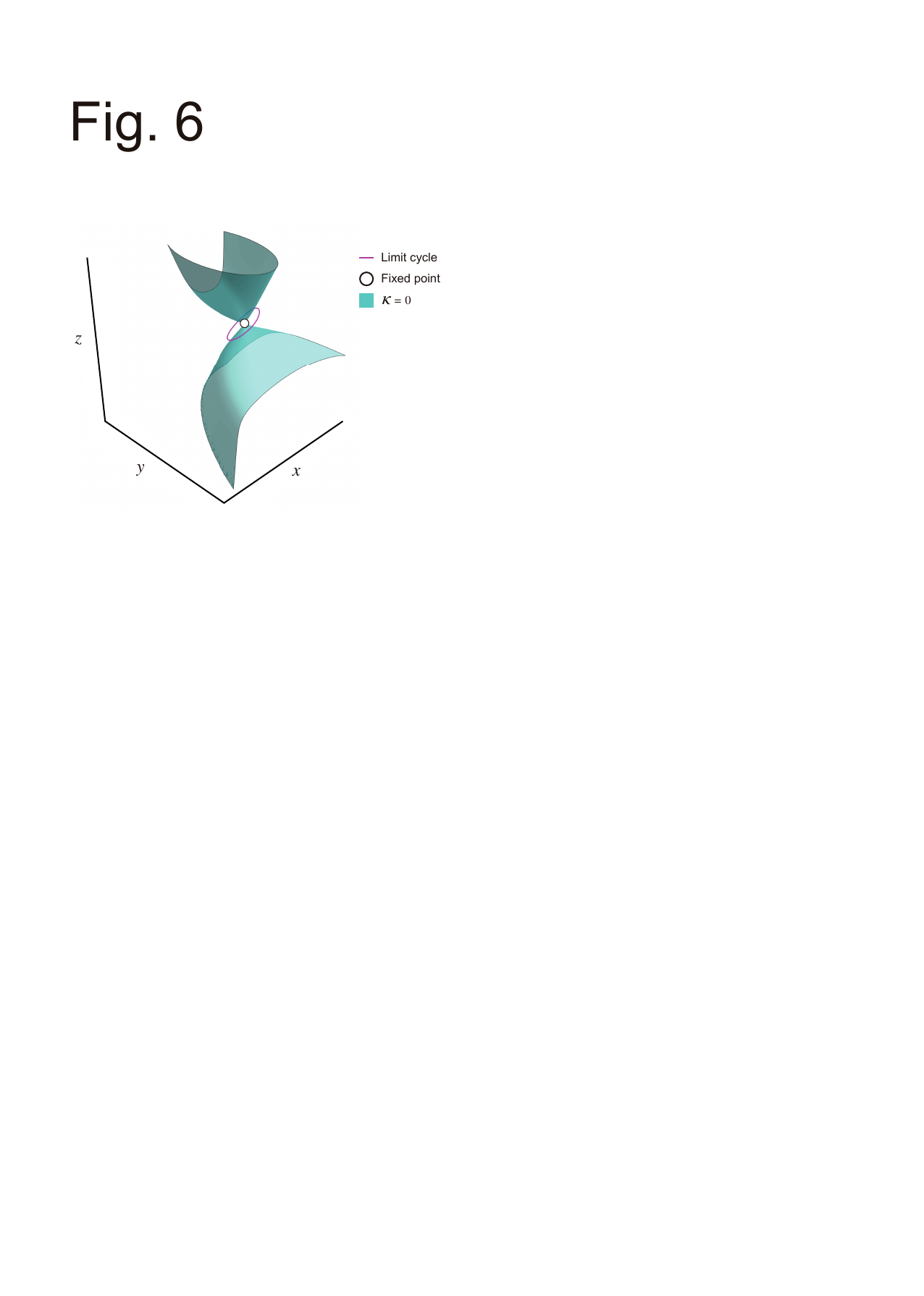}
\includepdf[pages=-]{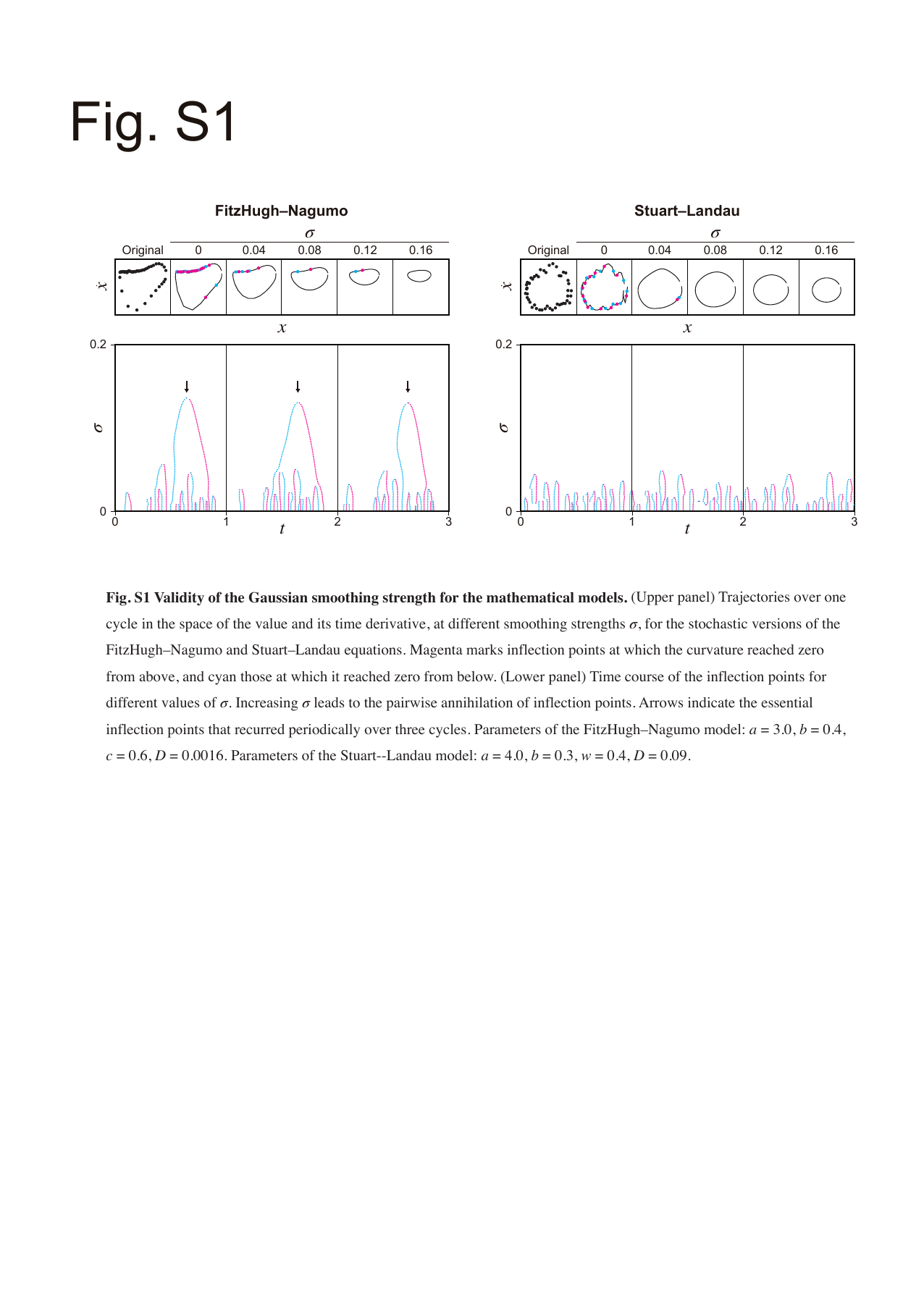}
\includepdf[pages=-]{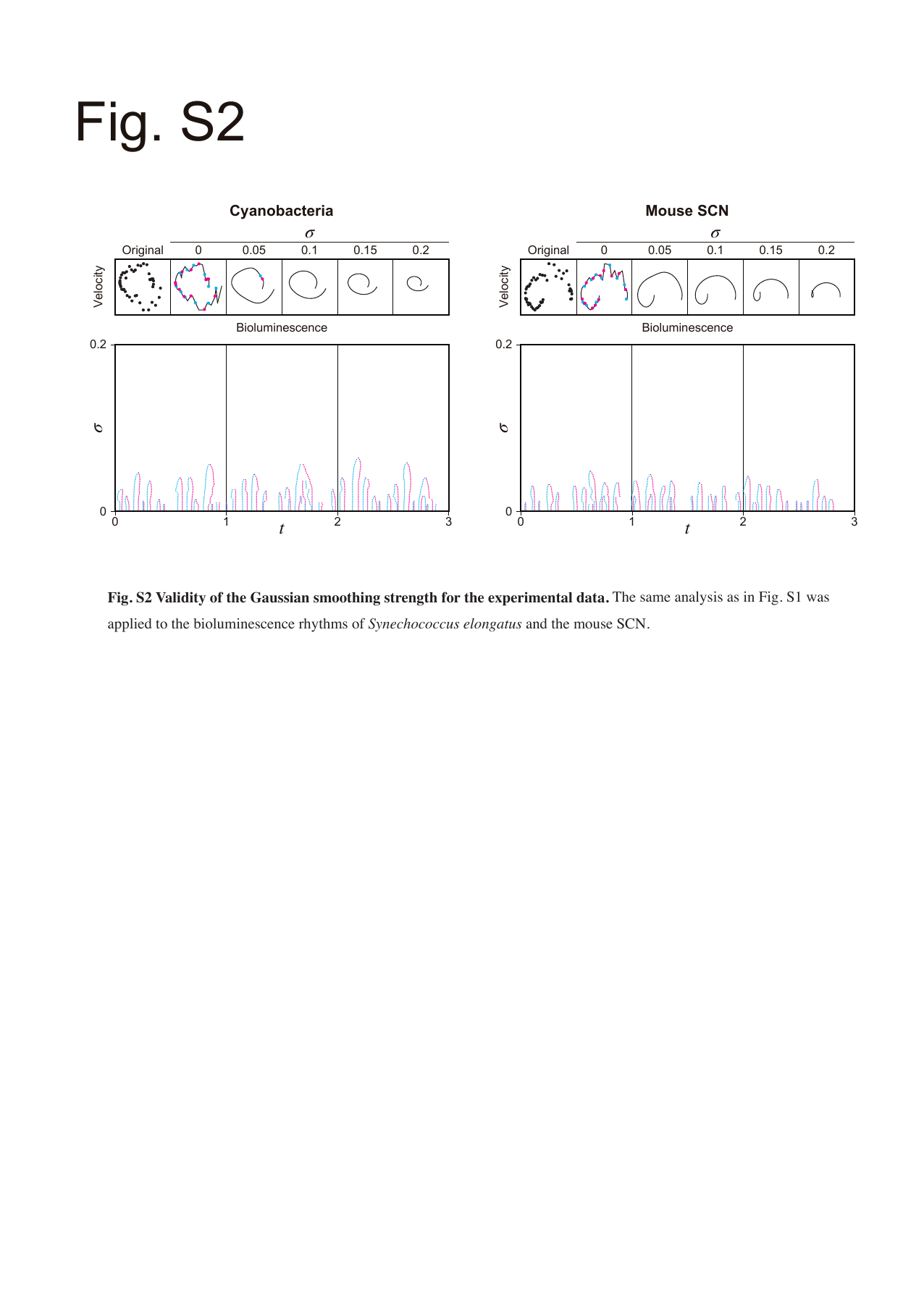}
\includepdf[pages=-]{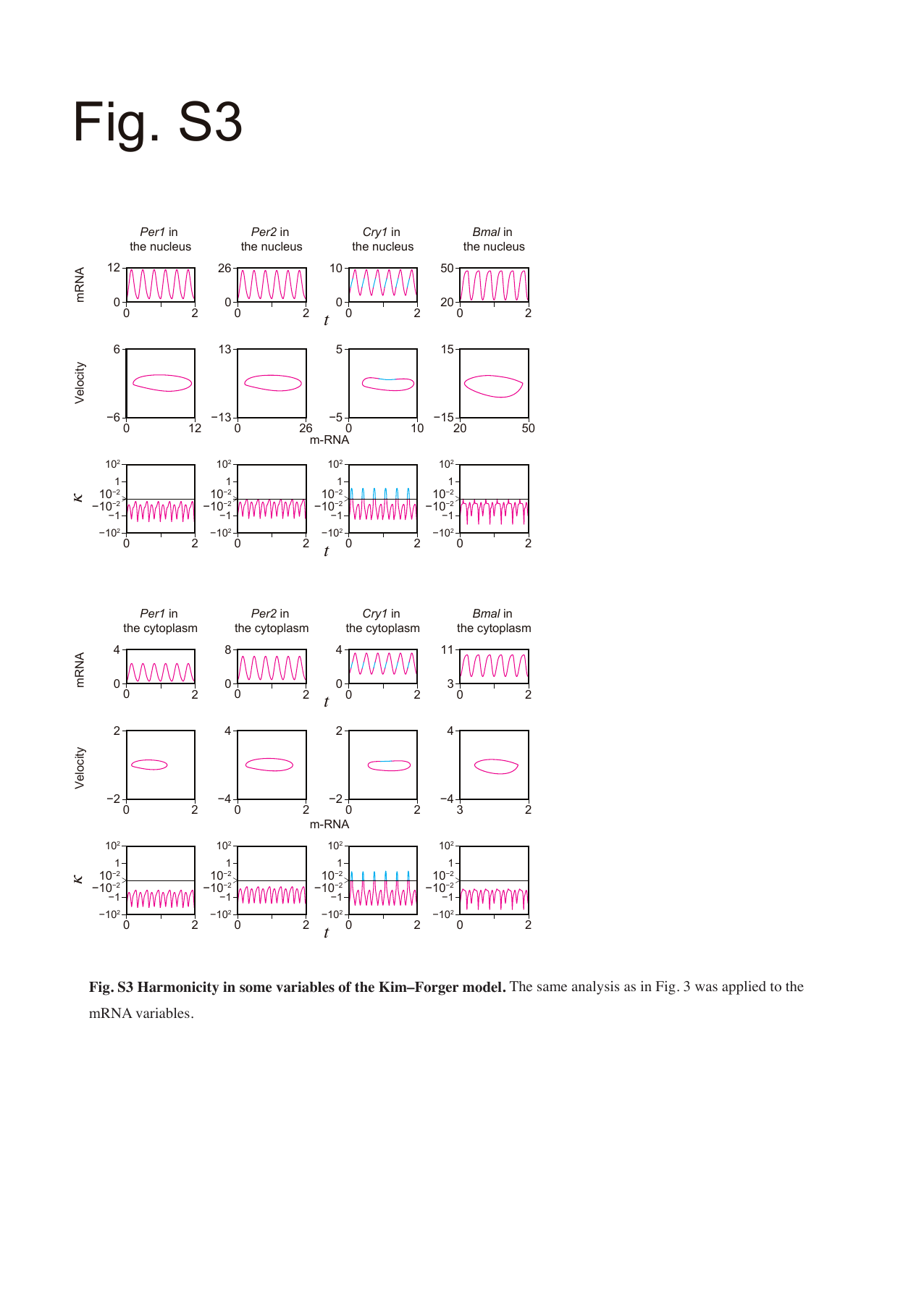}

\end{document}